\documentclass[twocolumn]{aastex62}

\usepackage[utf8]{inputenc}
\DeclareUnicodeCharacter{2212}{-}
\usepackage{graphicx}
\graphicspath{{./}{./figures/}{./tex/figures/}}
\usepackage{amsmath}
\usepackage{afterpage}
\usepackage{enumitem}
\usepackage{xcolor}
\usepackage{float}
\setenumerate{itemsep=0mm}

\usepackage{lineno}

\usepackage{soul} 
\usepackage{amsmath}
\usepackage{amssymb}
\usepackage{xspace}
\usepackage{xifthen}
\usepackage{eso-pic}

\newcommand{\Gaia}{{\it Gaia}\xspace}

\definecolor{forestgreen}{HTML}{228B22}
\definecolor{urlblue}{HTML}{000000}

\mathchardef\mhyphen="2D

\newcommand{\roughly}{\ensuremath{ {\sim}\,} }

\newlength{\dhatheight}

\newcommand{\code}[1]{\texttt{#1}\xspace}

\newcommand{\unit}[1]{\ensuremath{\mathrm{\,#1}}\xspace}

\newcommand{\degree}{\ensuremath{{}^{\circ}}\xspace}

\newcommand{\cm}{\unit{cm}}

\newcommand{\secref}[1]{Section~\ref{sec:#1}}

\newcommand{\tabref}[1]{Table~\ref{tab:#1}}

\newcommand{\figref}[1]{Figure~\ref{fig:#1}}

\newcommand{\bandvar}[2][]{%
  \ifthenelse{\isempty{#1}}{\var{#2}}{\var{#2\_#1}}%
}

\newcommand{\var}[1]{\ensuremath{\texttt{\MakeUppercase{#1}}}\xspace}

\providecommand\physrep{\ref@jnl{Phys.~Rep.}}%
\providecommand\apjs{\ref@jnl{ApJS}}%
\providecommand{\jcap}{\ref@jnl{JCAP}}%

\shorttitle{Photometric Study of a 1903 Photographic Plate}
\shortauthors{Yerkes Photographic Plate Digitization Team}

\begin{document}

\title{Precise Photometric Measurements from a 1903 Photographic Plate Using a Commercial Scanner}

\author[0000-0003-1697-7062]{William Cerny}
\affiliation{Department of Astronomy and Astrophysics, University of Chicago, \\ 5640 S. Ellis Ave., Chicago IL 60637, USA}
\affiliation{Kavli Institute for Cosmological Physics, University of Chicago, \\ 5640 S. Ellis Ave., Chicago, IL 60637, USA}

\author{Alexis Chapman}
\affiliation{Department of Astronomy and Astrophysics, University of Chicago, \\ 5640 S. Ellis Ave., Chicago IL 60637, USA}

\author{Rowen Glusman}
\affiliation{Department of Astronomy and Astrophysics, University of Chicago, \\ 5640 S. Ellis Ave., Chicago IL 60637, USA}

\author[0000-0003-2643-7924]{Richard G. Kron}
\affiliation{Department of Astronomy and Astrophysics, University of Chicago, \\ 5640 S. Ellis Ave., Chicago IL 60637, USA}

\author{Yingyi Liang}
\affiliation{Department of Astronomy and Astrophysics, University of Chicago, \\ 5640 S. Ellis Ave., Chicago IL 60637, USA}

\author[0000-0003-1266-3445]{Jason J. Lin}
\affiliation{Department of Astronomy and Astrophysics, University of Chicago, \\ 5640 S. Ellis Ave., Chicago IL 60637, USA}

\author[0000-0002-8397-8412]{Michael N. Martinez}
\affiliation{Department of Astronomy and Astrophysics, University of Chicago, \\ 5640 S. Ellis Ave., Chicago IL 60637, USA}

\author{Elisabeth Medina}
\affiliation{Department of Astronomy and Astrophysics, University of Chicago, \\ 5640 S. Ellis Ave., Chicago IL 60637, USA}

\author{Amanda Muratore}
\affiliation{Department of Astronomy and Astrophysics, University of Chicago, \\ 5640 S. Ellis Ave., Chicago IL 60637, USA}

\author{Buduka Ogonor}
\affiliation{Department of Astronomy and Astrophysics, University of Chicago, \\ 5640 S. Ellis Ave., Chicago IL 60637, USA}

\author[0000-0002-9142-6378]{Jorge A. Sanchez}
\affiliation{Department of Astronomy and Astrophysics, University of Chicago, \\ 5640 S. Ellis Ave., Chicago IL 60637, USA}

\collaboration{(Yerkes Photographic Plate Digitization Team)}
\correspondingauthor{William Cerny}
\email{williamcerny@uchicago.edu}

\begin{abstract}
We demonstrate the feasibility of determining magnitudes of stars on archival photographic plates using a commercially available scanner. We describe one photometric approach that could serve as a useful example for other studies. In particular, we measure and calibrate stellar magnitudes from a 1903 photographic plate from the Yerkes Observatory collection, and demonstrate that the overall precision from our methods is better than 0.10 mag. Notably, these measurements are dominated by intrinsic plate noise, rather than noise introduced through the scanning/digitization process. The low expense of this approach expands the scientific potential to study variable stars in the archives of observatory plate collections. We use the serendipitous discovery of a candidate transient at photographic magnitude $pg$ = 16.60 in the spiral galaxy NGC~7331 to illustrate our photometric methods. If  this unknown source is a supernova, it would represent the fourth known supernova in NGC~7331. 

\end{abstract}

\keywords{photographic photometry, supernovae}


\section{Introduction}
\label{sec:intro}

Beginning in the 1880s, the development of emulsion-coated (dry) glass photographic plates offered astronomers the unprecedented sensitivity and stability necessary to record long-exposure observations of the sky, revolutionizing astronomy.  Observatories shortly began to amass substantial collections of plates: by 1890, Harvard College Observatory alone had used photographic plates to collect and classify the spectra of more than 10,000 stars - a number which increased to more than 250,000 spectra by the mid 1920s. While charge-coupled devices (CCDs) came into use in astronomy in the late 1970's, the use of photographic plates continued for wide-field applications well into the 1990's. 
\par Although CCD detectors proved to be far more sensitive and are now ubiquitous in astronomical science, the plate collections retained by worldwide observatories still hold significant value. In particular, the study of the photometric and astrometric data contained on astronomical photographic plates offers a unique opportunity to probe astrophysical processes, trace orbits, and analyze variation over a timescale much longer than the $\roughly 30$ years in which CCD technology has been widely available.
The annotations and metadata (e.g, information including the instrument used, type of photographic plates, exposure times, etc) associated with these plates and their corresponding observatory logs are also interesting in their own right, as they trace the history and methodology of astronomy throughout the late 19th and 20th centuries.
\par Despite their clear historical and scientific value, the sheer number of astronomical plates in observatory collections ($>2.4$ million in the US and Canada alone; \citealt{census}) combined with the difficulties associated with their preservation has sparked considerable discussion within professional astronomical organizations (including the AAS and IAU) about the most practical approaches to extract the relevant information. 
Some past efforts have stressed the historical value of collecting and providing public digital access to plate metadata and observing logs. Compared with the overall size of the collections, relatively few photographic plates have been digitized in a suitable manner for precision photometric and astrometric measurement.
Large-scale efforts to digitize photographic plates have included the creation of the Hubble Space Telescope Guide Star Catalog \citep{Lasker}, the Automatic Plate Measuring machine (APM; \citealt{APM}) and COSMOS \citep{cosmos} programs in the United Kingdom, and the Digital Access to a Sky Century at Harvard (DASCH; \citealt{dasch}) project. 
Most recently, the (ongoing) DASCH project has developed a high-speed digital plate scanner and analysis pipeline capable of efficiently and homogeneously digitizing and processing plate data. However, while the DASCH project has digitized more than 400,000 plates to date, the investment required renders similar campaigns for other institutional collections unlikely. Therefore, there is considerable incentive to explore the potential of commercially available digitizers to undertake complementary efforts.
\par A number of earlier investigations have attempted to use flatbed scanners to digitize astronomical plates, many of which have been concerned with astrometric measurements of stars  \citep[e.g.,][]{Vicente07}. Other studies have been primarily concerned with photometric measurements (\citealt{Lamareille, Pakuliak, Sokolovsky}). These photometry-focused studies have typically reported internal photometric precision of order 0.1 mag --- a result that provides context for our analysis.  For comparison, special digitizers built for the purpose of making photometric measurements on glass plates yield similar levels of precision (\citealt{Ortiz-Gil,Rapaport, Grindlay14}), albeit in the case of DASCH with much greater digitization speed. 
\par In this work, we present the first results from our efforts to digitize a sample of photographic plates from the Yerkes Observatory plate collection. In \secref{selection}, we introduce the Ritchey 24" reflector and the photographic plate we selected for illustration. In \secref{digitize}, we describe the commercially available scanner we utilized to digitize this plate, discuss some of the ways we have characterized the performance of the scanner, and present our methods for converting the scanner output into a scientifically useful (calibrated) FITS file. In \secref{conversion}, we use the results of the transformations from \secref{digitize} and compare surface photometry between our plate and an image of the same field from the Sloan Digital Sky Survey. \secref{astrometry} quantifies the astrometric performance of the scanner and \secref{photometry} quantifies the overall photometric precision that we have achieved.  \secref{guest} gives a specific example of measuring a star that is not in a catalog; and in \secref{conclusion}, we summarize the main results concerning the astrometric and photometric precision we were able to achieve. 
\section{Yerkes Plate Collection and Plate Selection}
\label{sec:selection}

The Yerkes Observatory is home to between 150,000 and 200,000 photographic plate images and spectra, the vast majority of which were taken either at Yerkes itself or at the McDonald Observatory. The plates in the collection come from telescopes of widely varying focal length, design, and purpose, each with its own peculiarities. We chose to limit this analysis by using plates from only one telescope, the 24-inch Ritchey reflector.
This telescope, which was in service at Yerkes from 1901 until the 1960s when it was moved to the Smithsonian, was one of the first reflecting telescopes used successfully by professional astronomers and spurred the adoption of reflecting telescopes at Mt. Wilson and elsewhere.  Ritchey's telescope \citep{ritchey_tel} is a classic Newtonian design (parabolic primary and flat secondary) with a 93-inch focal length.  At full aperture (23.5 inches), the focal ratio is f/4. In addition to George Ritchey himself, the 24-inch was notably used by Edwin Hubble for his doctoral thesis (\citealt{poster}).
The reflector's plates are relatively small (usually 3.25 by 4.25 inches) and therefore easy to handle.
\par Using the telescope's logbooks, we selected and retrieved plates based on three criteria: exposure time, sky conditions, and sky location.
We chose plates with longer exposure times (usually over one hour) that would give an idea of the telescope's limiting magnitude.
Comments in the telescope logbook allowed us to select plates with good atmospheric conditions and avoid nights with a full moon, clouds, or bad seeing. We chose fields outside of the Galactic plane so that we could clearly measure extragalactic objects as well as stars.

\par After retrieving candidate plates, we visually inspected each one to look for defects such as cracks that could make scanning and analysis difficult. While many of our plates were annotated, the annotations generally did not interfere with the measurement process. We also selected plates without artifacts caused by poor photographic technique such as adjacency effects (e.g. rings of lower photographic density around bright stars). Over 50 plates were found that fulfilled these criteria.\footnote{A separate study by our team, \citet{AAS1}, used 12 of these plates to study the variability of field quasars.} Upon further tests, we found that plate Ry60 centered at (R.A, Decl.) = ($339.38\degree, +34.38\degree$) has an especially high star count because of its superior depth and its relatively low Galactic latitude ($b = -21\degree$). The large number of stars (over 7000) enables us to partition the sample and maintain good statistics for astrometric and photometric analysis, as described in \secref{astrometry} and \secref{photometry}.
\begin{figure}
    \centering
    \includegraphics{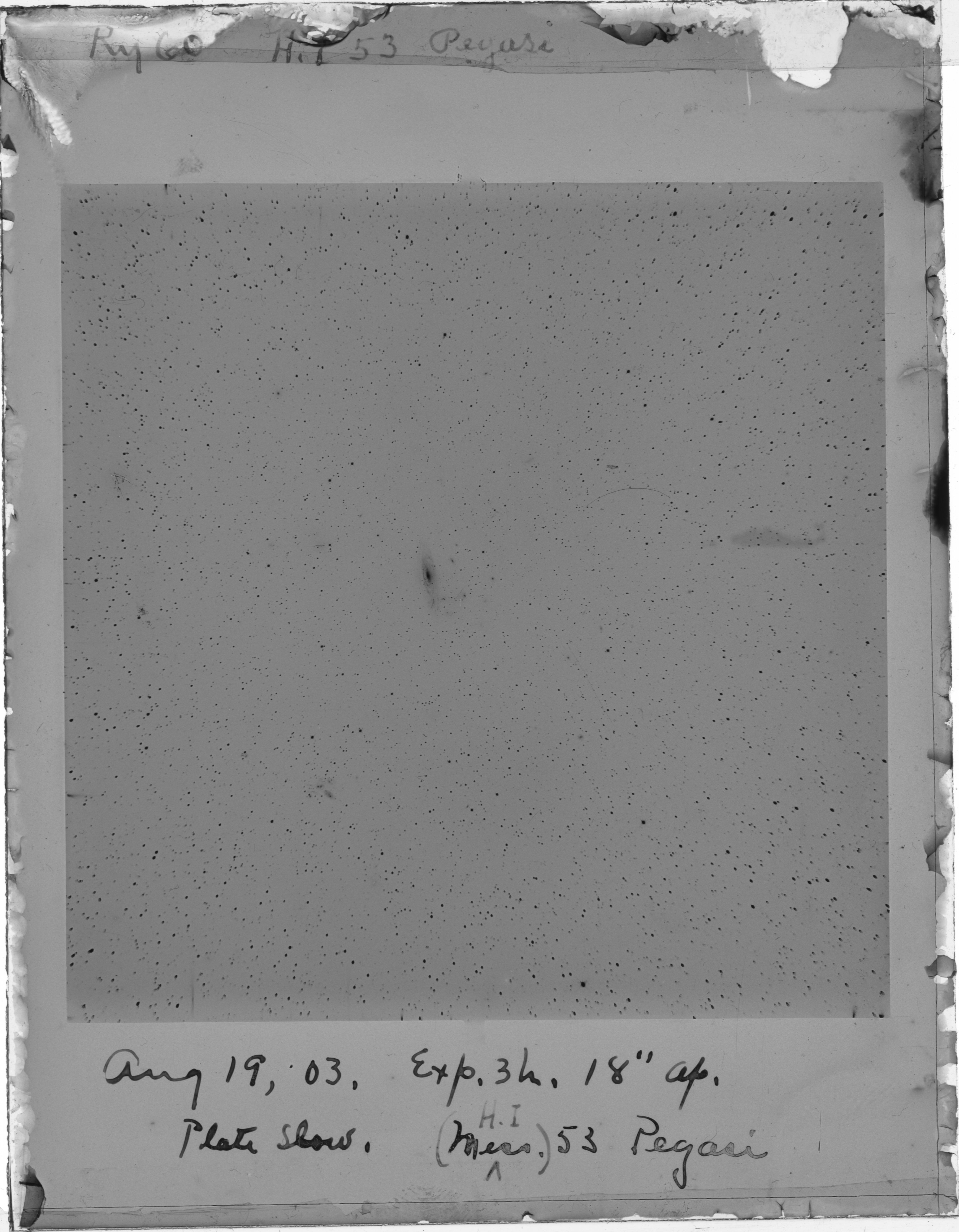}
    \caption{Plate Ry60, taken in 1903, dimension 3.25 x 4.25 inches. The field is a bit less than two degrees on a side.  The object in the center is NGC~7331; other members of the NGC~7331 Group are also present on the plate, as is Stephan's Quintet, which is between NGC~7331 and the lower-left corner.  H.I 53 refers to W. Herschel's original designation for NGC~7331.}
    \label{fig:plate}
\end{figure}
\par Ry60, shown in \figref{plate}, contains a large spiral galaxy, NGC~7331, that is useful for tests of scanner performance on a diffuse source, which we undertake in \secref{conversion}.  Moreover, the plate has a relatively uniform sky background that removes one potential source of photometric error.  Ry60 was exposed by George Ritchey on the night of August 19, 1903, for three hours, with the telescope aperture stopped down to 18 inches (a technique that lessens the amount of coma).  We do not have information on the emulsion, but Seed~27, a blue-sensitive emulsion, was used at Yerkes Observatory in this time period. In \secref{photometry}, we demonstrate that our derived color term also indicates a blue bandpass with an effective wavelength near 420 nm. 

Our broad approach in this paper is to test measurements derived from digitized scans of Ry60 against astrometry and photometry from the Sloan Digital Sky Survey Data Release 7 (SDSS DR7; \citealt{sdss,sdssDR7}) and \Gaia Data Release 2 (DR2; \citealt{gaiaDR2}).  Residuals may result from a number of sources:  i) intrinsic plate noise; ii) errors in the World Coordinate System (WCS) solution for the plate; iii) noise introduced by the digitizer; iv) image blending; and v) stellar variability in position or flux. As described in \secref{digitize}, we contrived some deliberate tests of the digitizer to evaluate its performance, but in general the net residuals provide an end-to-end demonstration of the precision we can achieve, folding all the sources of error together. 
\par In \tabref{summary}, we summarize key properties of the Ry60 plate studied in this work.
\begin{deluxetable}{l c}[h]
\tablecolumns{2}
\tabletypesize{\small}
\tablecaption{\label{tab:summary}
Summary of Plate Properties}
\tablehead{
\colhead{Parameter}  & \colhead{Value}}
\startdata
Physical Dimensions & $3.25 \text{ in} \times   4.25 \text{ in} $ \\
Date of Exposure & August 19, 1903 \\
Exposure Time & 3 hours\\ 
Center Coordinates ($\alpha$,$\delta$)& ($339.38\degree, +34.38\degree$) \\
Center Coordinates ($\ell$,$b$)& ($93.79\degree, -20.80\degree$) \\
Sky Area Scanned &  $1.48\degree \times 1.48\degree $ \\
Physical Plate Scale & 87.4 arcsec/mm \\
Matched Sources with \Gaia & $\sim 6200$ \\
Limiting Magnitude (AB system) & $g \lesssim 19$ \\
\enddata
\vspace{-3em}
\end{deluxetable}

\section{Digitization Using a Commercial Scanner}
\label{sec:digitize}

We used a flatbed scanner with provision to scan transparencies, specifically the Epson Expression 12000XL graphic arts scanner.  The high-resolution performance of a similar model was discussed by \citet{simcoe}.  We configured the output of the scanner to be a 16-bit grayscale TIFF file with other options turned off.  We scanned the plate as a positive, meaning that the TIFF files show dark stars and a light background (\figref{plate}). We typically use a dots-per-inch (dpi) setting of 1600, lower than the resolution investigated by Simcoe.  At dpi = 1600, the step between samples is 15.9 microns.  Since the plate scale of the Ritchey reflector is 87.4 arcseconds/mm, a step is 1.39 arcseconds. Our scan was centered in the field of the plate and was 2.4 inches on a side, corresponding to 3840 pixels or about 1.5 degrees.
The image quality near the center of the field on Ry60 has a full-width at half-maximum (FWHM) of around six arcseconds, such that we sample each star image by several steps in each dimension.  
\begin{figure}
    \centering
    \includegraphics[width=\columnwidth]{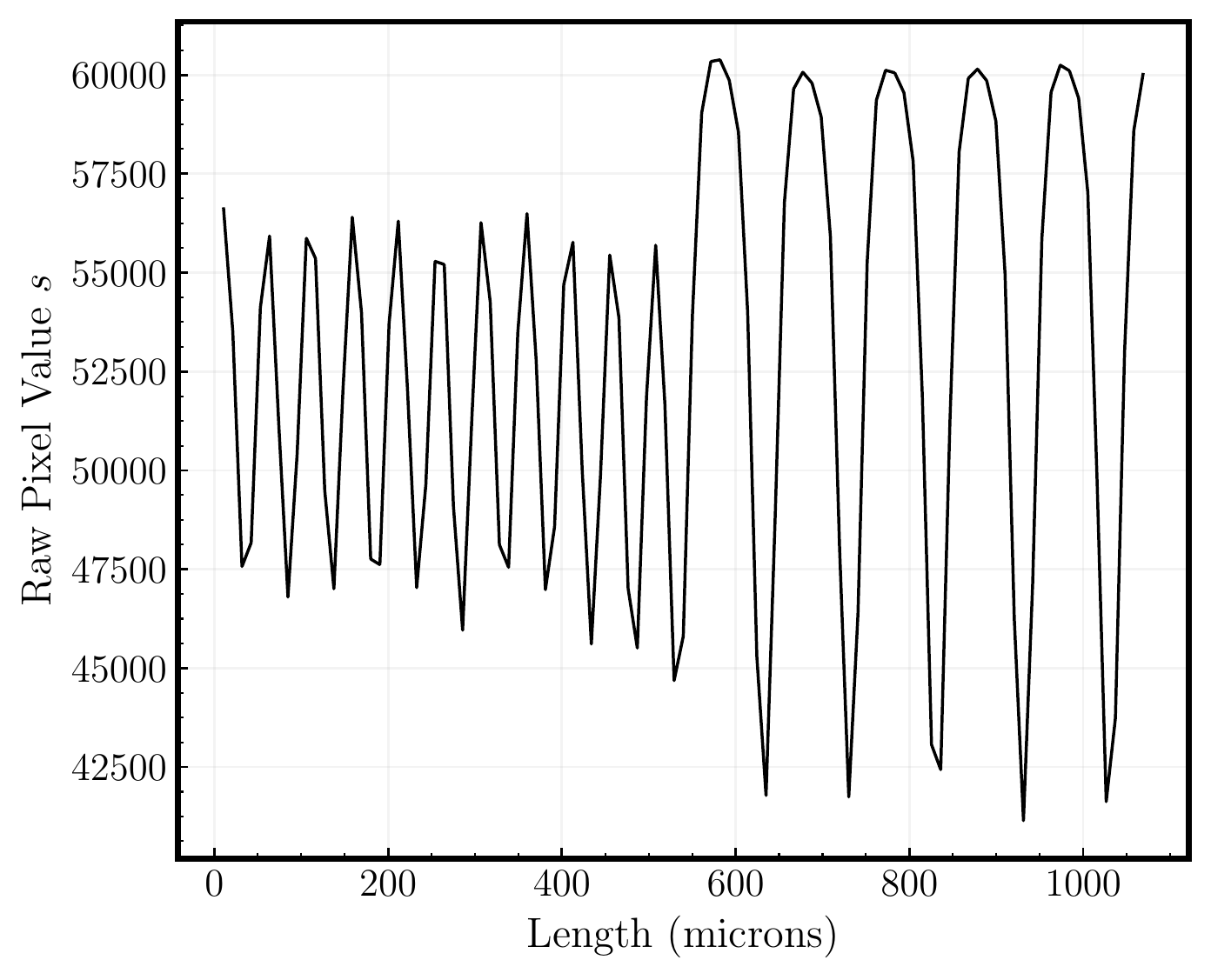}
    \caption{Test of the response of the scanner (raw pixel value) to narrow, closely spaced lines on the r\'eseau.  The lines on the left-hand side are spaced by 50 microns, or about 5 samples at the scanner setting of 2400 dpi.}
    \label{fig:reseau}
\end{figure}
\par We first implemented a test to explore the response of the scanner to edges, i.e. changes from dark to bright on the scale of a step size.  Neighboring pixels may be expected to be correlated at some level, scattered light being one physical cause. To analyze this effect, we scanned a r\'eseau with narrow, opaque lines with known physical line spacings of 50 microns and 100 microns.  \par \figref{reseau} shows the results of a scan of this r\'eseau, where the 50 micron spacing is on the lefthand side and the 100 micron spacing is on the righthand side. (Specifically, \figref{reseau} shows the scanner output when this pattern is scanned parallel to the lines; the perpendicular direction is only slightly worse.) For reference, the sharpest star images are near the center of the plate; these images have intrinsically shallower gradients but we also sample them more coarsely (1600 dpi for the plate and 2400 dpi for the r\'eseau). \figref{reseau} shows qualitatively that the scanner is capable of measuring features at a scale significantly smaller than 50 microns. 
\par
\begin{figure}
    \centering
    \includegraphics[width=\columnwidth]{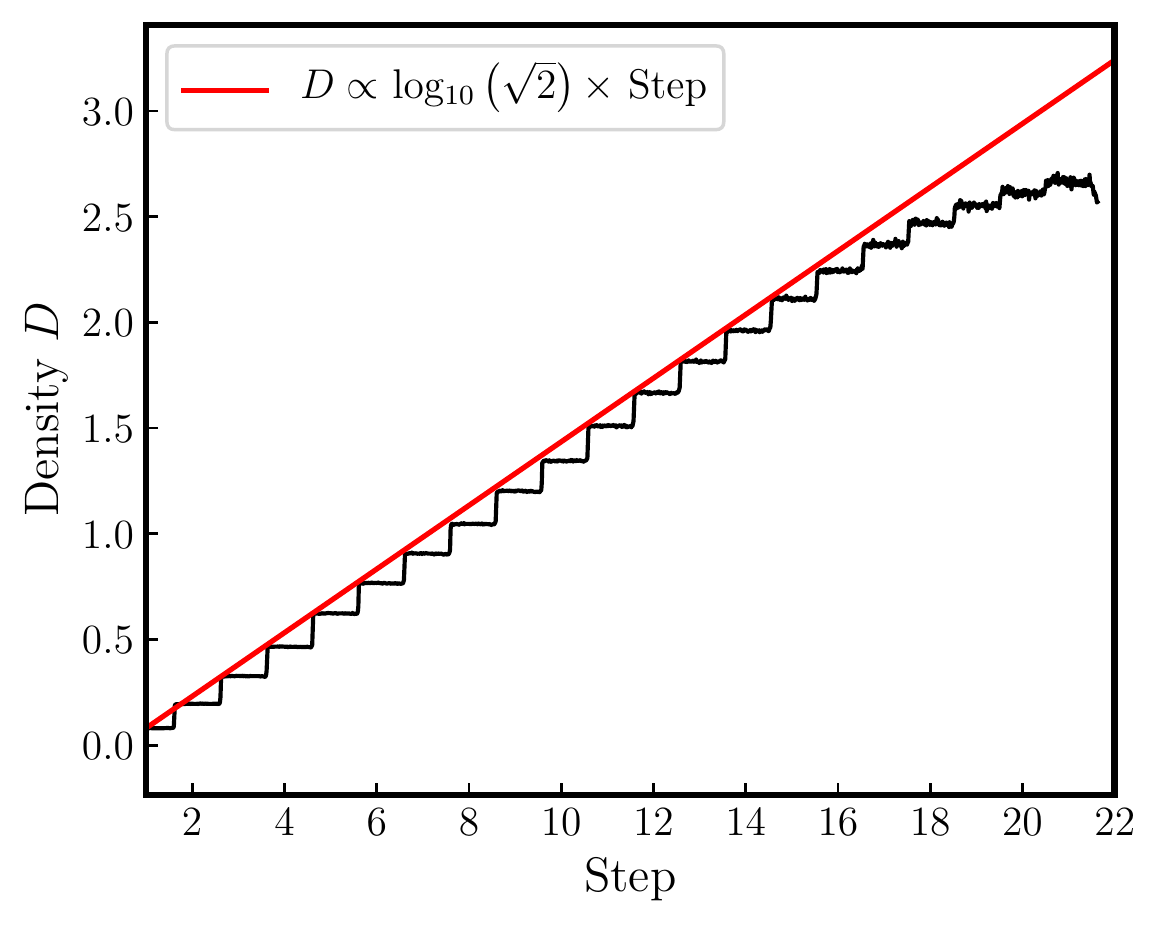}
    \caption{Scan of a calibrated 21-step wedge with steps (x-axis) incremented by $\log(\sqrt{2})$ in density. The wedge was scanned with same default parameters as for the plate. The values on the y-axis transform the raw scanner output with a two-parameter function as described in the text.}
    \label{fig:steps}
\end{figure}

 Another test explores the relationship between the pixel values produced by the scanner and photographic densities. The scanner values have some relationship to the transmission, or density, at each place on the plate.  To determine this relationship, we scanned a step wedge calibrated in terms of density. \figref{steps} depicts the relation between the input (x-axis) and the scanner measurement (y-axis), transformed as described below. Each step increments density by 0.15, corresponding to a factor of $\sqrt{2}$ in transmission. \figref{steps} shows that the highest densities are not measured well, but overall there is a good correspondence. 
Based on the scanner values and the step wedge densities, our prescription of photographic density values $D$ is:
$$ D = 1.4 \times \log_{10}{[65535/(s - 1900)]}$$
where $s$ is the original scanner output value for a pixel. The value 1900 can be considered as a correction for scattered light, and the value 1.4 can be considered as a scale factor that adjusts the raw scanner output numbers to true measurements of transmission. 
The transformed values that were entered into a new version of the file are proportional to  \(10^D\). We note that the sky level on the plate corresponds to photographic density $D = 0.43$. 
\par For further analysis, the original scanner TIFF file for plate Ry60 was converted to a FITS file with a WCS astrometric solution by uploading to \code{Astrometry.net} \citep{astrometry.net}.  Uploading the resulting FITS file back into the site again was found to give a larger number of matches of the detected sources to the index catalog and an improved WCS solution. We can then utilize these derived WCS coordinates to compare our plate to other astronomical images and catalogs, as we undertake throughout the rest of this work.

\section{Surface Photometry}
\label{sec:conversion}

\begin{figure*}
    \centering
    \includegraphics[width=500pt]{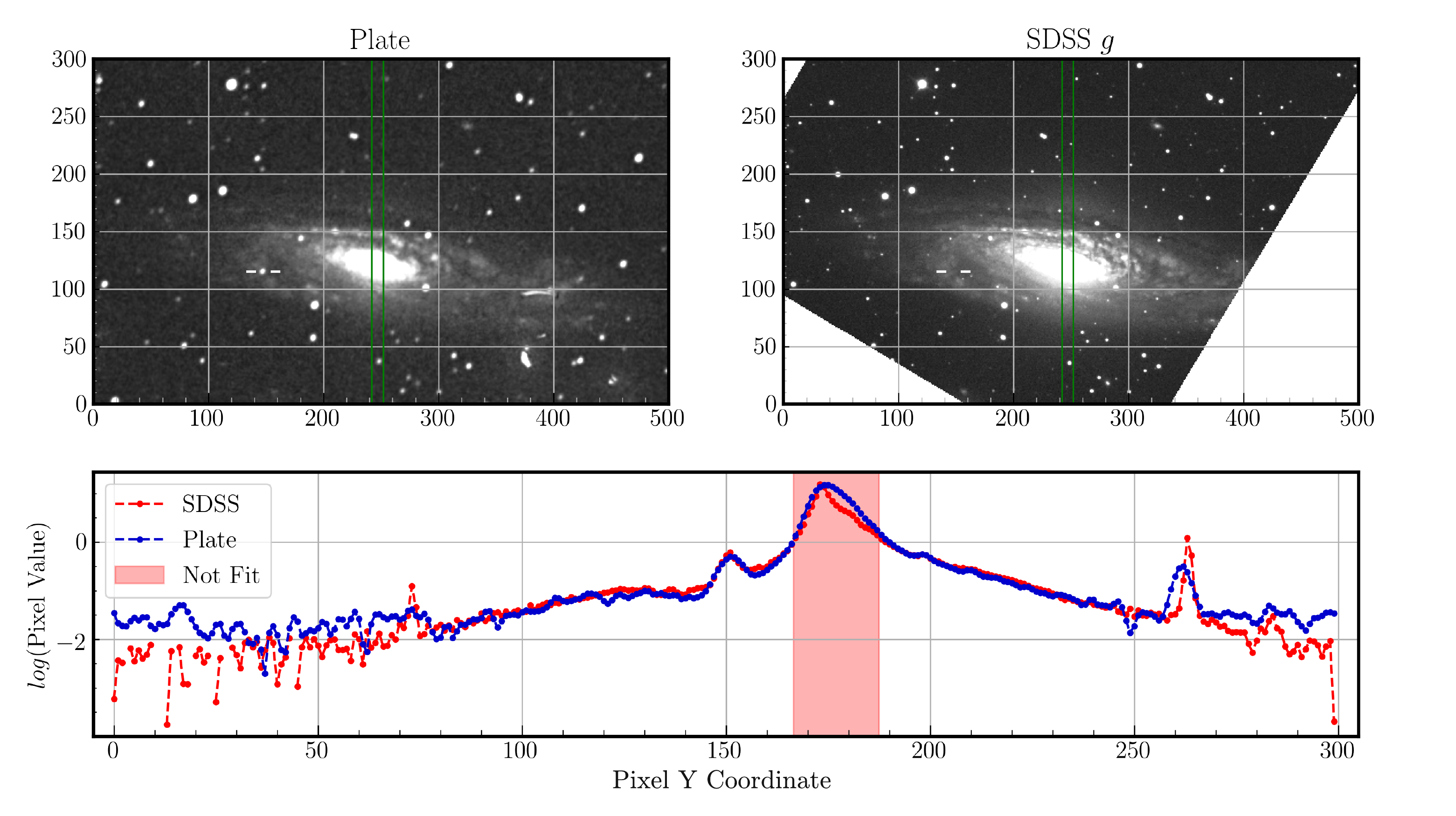}
    \caption{Comparison of SDSS $g-$band surface photometry with reprojected scanner values on plate Ry60. The lines in the lower graph represent the average of each row of pixels between the green lines in the upper images, where left to right in the graph corresponds to top to bottom in the images. East is at the top and North is to the right in the images. The red shaded area corresponds to points that were not considered when fitting the sky background and scaling parameters due to saturation on the plate. The white dash marks in both images correspond to the position of the `guest star' mentioned in \secref{guest}. The SDSS image corresponds to one SDSS frame; positions outside of the frame are shown in white.}
    \label{fig:n7331}
\end{figure*}

To see how well the transformation described in the previous section works in practice, in \figref{n7331} we compare the transformed values with SDSS $g-$band surface photometry for a selected region within the image of NGC~7331.   Besides the choice of the region, the only parameters in the comparison are the sky background to subtract from the plate values, and a fixed scale factor to convert from the sky-subtracted plate values to the SDSS $g-$band intensities. After obtaining the transformed pixels and reprojecting them to the SDSS image's scale, we fit a 2-parameter background and scaling model to match surface brightness values across the area. 
\par The mismatches in the two curves in \figref{n7331} inform a few conclusions about our transformation process. The biggest differences are present to the far left and right of the central peak, which sample the background at very low surface brightness. While the plate background could be adjusted down to better match the SDSS background, we found this adjustment introduced a larger mismatch at the brighter areas of the image. The other two obvious differences present in \figref{n7331} are the unmatched dip in SDSS values on the right of the central peak, and the spike in surface brightness to the right of the galaxy. The former corresponds to a dust lane in NGC~7331 that occurs on the plate at high densities in a regime that is difficult to calibrate accurately. The latter is a star, which appears thinner in the SDSS due to differences in point-spread function, and which appears offset in position due to its proper motion over the past century. Point sources provide significant challenges to our methods, but in regions with smoothly changing surface brightness, such as the galaxy itself, our transformation process was quite successful. As illustrated in \figref{n7331}, we see good agreement over a factor of 100 in surface brightness, covering essentially the entire visible range of NGC~7331 on the plate.  We conclude that the characteristic curve (density $D$ vs. log intensity) for the plate is approximately linear over this range of intensity. 

\begin{figure*}
    \centering
    \includegraphics[width=300pt]{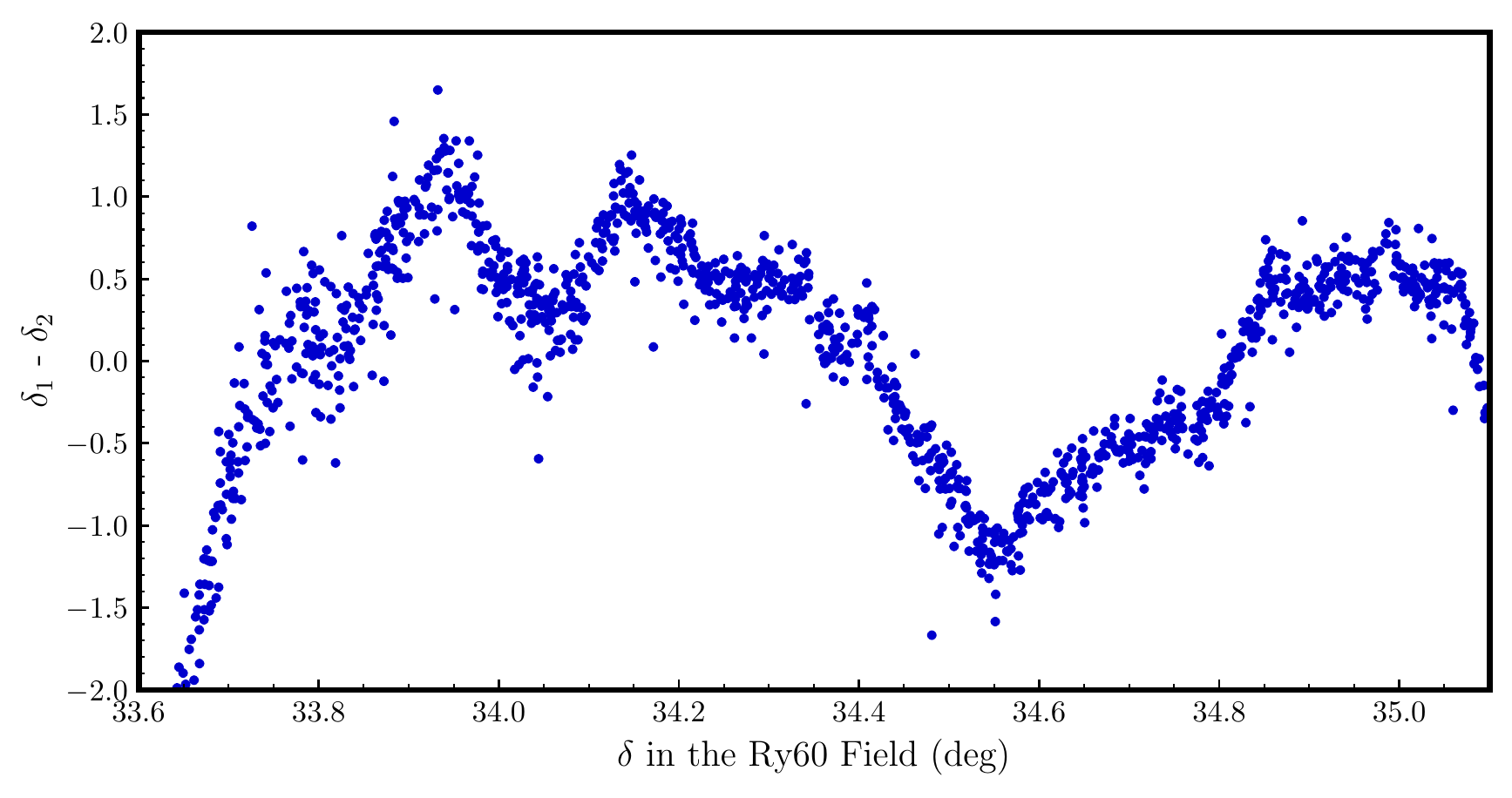}
    \caption{Excursions found in our measurements of positions, here in decl., when plotting the difference between scans rotated by 90 degrees. The excursions have an amplitude of around $\pm 1.5$ arcseconds. At a given declination, the rms scatter is about $\pm 0.12$ arcseconds. }
    \label{fig:del_dec}
\end{figure*}

\begin{figure*}
    \centering
    \includegraphics[width=400pt]{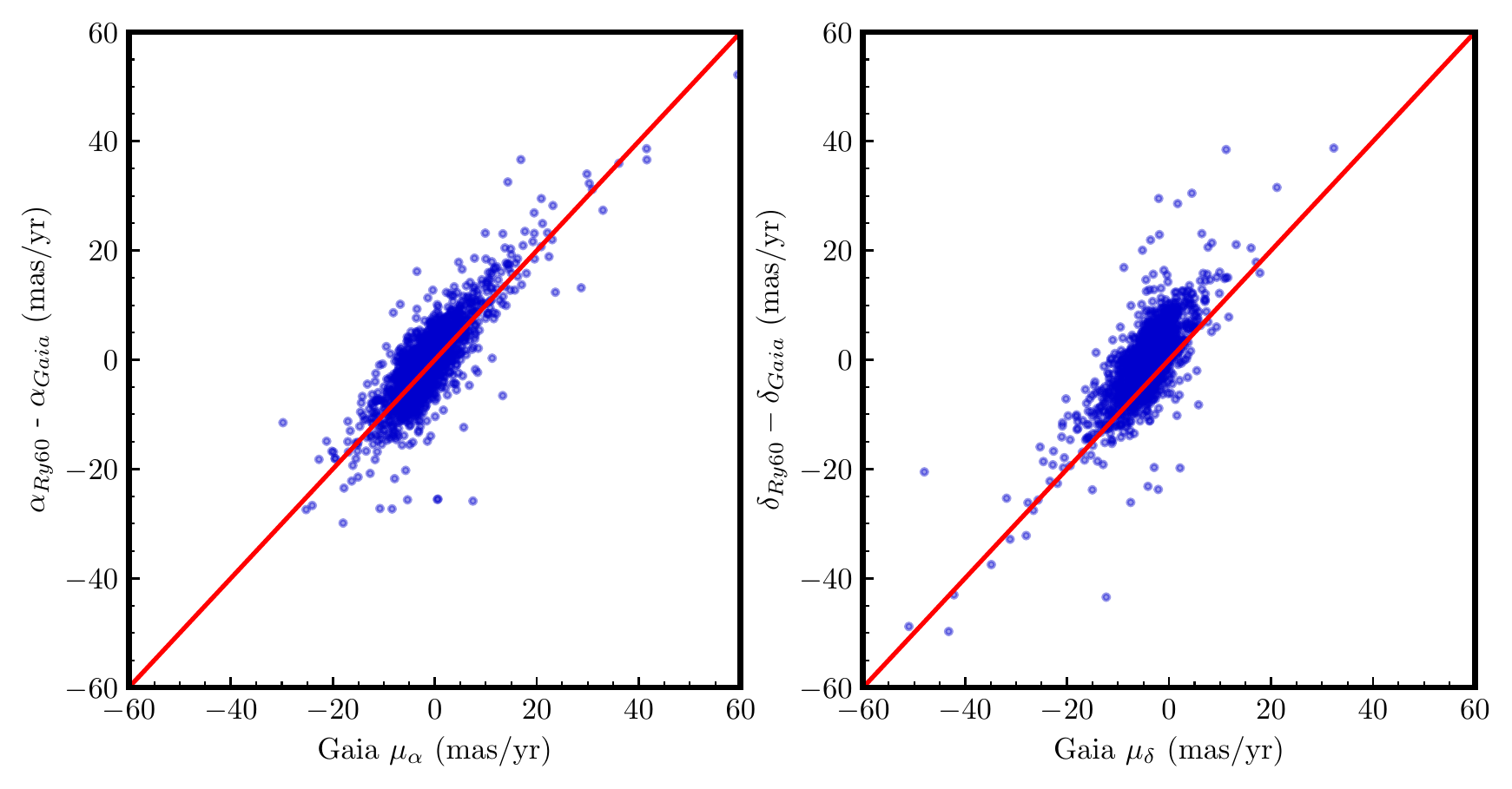}
    \caption{Proper motion in both Right Ascension and Declination from the \Gaia DR2 catalog (x-axis) against the difference between the position of the average of the perpendicular scans of Ry60 and \Gaia. This difference is then scaled to units of milliarcseconds/year by accounting for the 112-year baseline (y-axis). The sources plotted are less than 29 arcminutes from the center of the plate and have a \Gaia DR2 BP magnitude between 15.5 and 18.}
    \label{fig:PM}
\end{figure*}

\section{Stellar Catalog and Astrometry}
\label{sec:astrometry}
In this section, we describe our procedure for source detection, automatic photometric measurement, and astrometric measurements. Our approach focuses on optimizing global residuals and identifying/correcting trends over several thousand stars detected on the plate, as opposed to optimizing the measurement of an individual star.
\subsection{Object Detection and Measurement}
To detect sources in the Ry60 field and to obtain (R.A., Decl.) centroids and sky-subtracted fixed-aperture measures of brightness, we use the \code{Aperture Photometry Tool} (APT; \citealt{APT}). APT's source-detection routine requires the user to specify some parameters including the minimum and maximum number of pixels expected for a given source and the minimum sigma above background for a source to be detected confidently.  APT's photometric measurement then follows based on the specified aperture for measuring each source and the specified region for evaluating the sky background. For illustration in this work, we used a measurement aperture of radius four pixels ($\roughly5.6$ arcseconds), and the sky background was determined locally by the median value for all pixels between 20 and 30 pixels distance from the centroid of each star. With these parameters specified, sources in the field are automatically detected, and in a subsequent pass are automatically measured.  The output file consists of the positions of the stars along with the aperture measurement and several other quantities. 
\par For sake of analyzing the reproducibility of our results, we rotated the Ry60 plate by $90\degree$ and made another TIFF file and associated FITS file following the procedure described in \secref{conversion}. This analysis was motivated by non-uniformity in the effective scanner steps in the scan direction, resulting in astrometric variation, discussed in detail below.
\subsection{Astrometric Performance}
\par We undertake two distinct tests to assess the  performance of the astrometry derived from the procedures described in \secref{digitize}.
\par Firstly, the astrometric reproducibility of the scanner can be assessed by comparing the star positions as measured on the two scans taken in perpendicular directions. We use the Tool for OPerations on Catalogues And Tables (\code{TOPCAT}; \citealt{topcat}) match utility to cross-match the source lists provided by APT along R.A. and decl. columns, resulting in an output list containing the $\sim 7600$ sources found in both scans. By construction, the scan direction was closely aligned with either R.A. or decl. Since the WCS solutions should be very similar for the two scans, errors due to an imperfect modeling of field distortion should cancel out, and the differences in positions should reflect the scan-direction error of the scanner. 
\par The difference in declination between the two scans is plotted against declination across the field of Ry60 in \figref{del_dec}. (The corresponding plot for Right Ascension looks qualitatively similar.) The behavior can be characterized by excursions in the data with an amplitude of around 1.5 arcseconds on a spatial scale of $\roughly1 \cm$, on top of which is smaller-amplitude scatter; over a small range in declination the scatter is $\pm 0.12$ arcseconds. Thus, if the large-scale excursions were modelled from a sufficiently dense network of stars, this value of the scatter would be the irreducible astrometric error. 

\par Secondly, we compare our stellar positions against the high precision astrometry provided by \Gaia DR2 in order to illustrate the external (absolute) astrometric precision by calculating the proper motion of sources over our time baseline and comparing to the catalogued values. To do so, we generate a source list from the same field as our plate image via a cone-search query to ESA's \Gaia DR2 catalog\footnote{Data from \Gaia DR2 is available at \url{https://gea.esac.esa.int/archive/}}. To reduce effects of the orientation of the scanning, we take the average of the positions from both scanned orientations, and crossmatch these positions, again with \code{TOPCAT}, to isolate all of the sources which lie both on the plate and exist within the \Gaia DR2 catalog.  Imposing a angular separation tolerance of six arcseconds, we find $\sim6200$ unique sources were matched between the averaged plate scans and \Gaia, with a median positional difference of 1.4 arcseconds, which corresponds to roughly one pixel.
\par In addition to R.A./decl. coordinate information, the \Gaia catalog also provides measurements of stellar proper motions in both R.A. and decl. ($\mu_{\alpha}$ and $\mu_{\delta}$, respectively). These proper motions can be compared with the apparent shift in stellar positions between the plate's exposure epoch and the \Gaia measurement epoch in order to assess whether our procedure is sufficiently precise to detect stellar proper motions.  To calculate proper motion via the position on our plate, we take the difference between the aforementioned averaged position and the position from the \Gaia query. Since the \Gaia DR2 coordinates are reported in the J2015 epoch and Ry60 was taken in 1903, we divide the difference by the 112 year baseline. The units in the position difference are then converted to milliarcseconds per year. 
\par In \figref{PM}, we compare these derived proper motions with those from the \Gaia catalog for the same sources. The relationship between the two confirms that our plates and methods are sufficient to detect accurate stellar proper motions over a baseline of 112 years.
\section{Photometric Calibration and Performance}
\label{sec:photometry}
In this section, we demonstrate the photometric performance of the scanner by utilizing the aperture measurements from the previous section to assess our level of photometric repeatability, and then utilize photometry from SDSS DR7 and \Gaia to estimate the plate's limiting magnitude and calibrate our photometric $mag$ measurements.

\subsection{Photometric Repeatability}
In order to quantify the degree of photometric reproducibility of our measurements, and assess the effects of the orientation of the plate on the flatbed of the scanner, we return to the consecutive scans of Ry60 that were taken at orientations differing by 90 degrees while holding all other variables fixed. The digitzation process is expected to introduce some noise to the underlying plate data in addition to slight positional shifts of star images between scans, which lead to the projected four-pixel-radius aperture used by APT encompassing slightly different pixels within each scan.
\par
\begin{figure}[h]
    \centering
    \includegraphics[width=250pt]{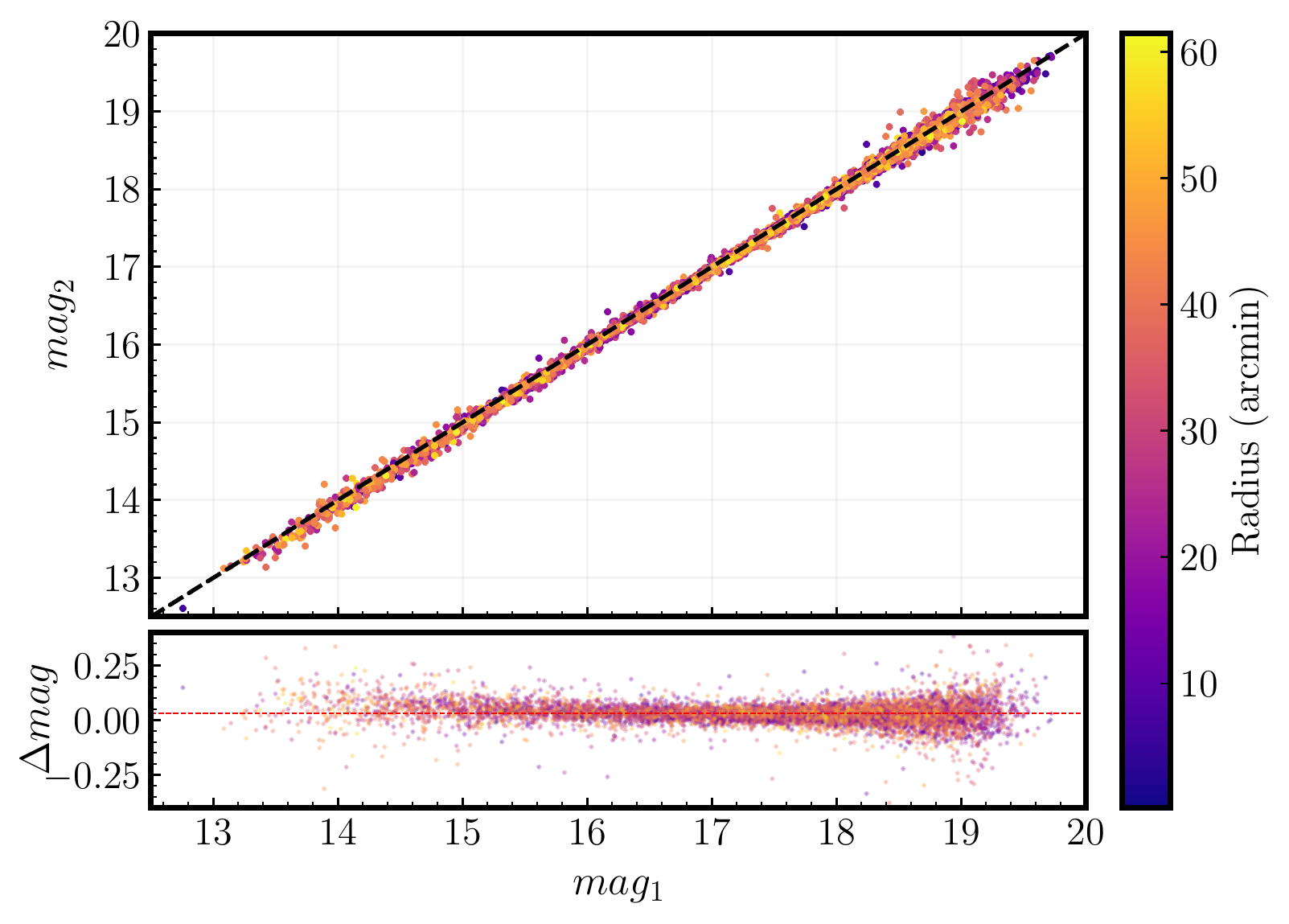}
    \caption{Comparison of $mag$ measurements from two scans of Ry60 rotated by 90 degrees as a function of radius, denoted by color. In general the magnitudes are very tightly correlated, indicating a reassuring level of reproducibility. No radial dependence is evident in the difference between the $mag$ measurements. There is some deviation from unit slope (black dashed line, top panel), and a one-sigma scatter about the mean of $\sim 0.03$ mag.}
    \label{fig:magvmag}
\end{figure}

 In \figref{magvmag}, we plot the correspondence between the two catalogs of APT-derived aperture magnitudes from the rotated scans, which we label as $mag_1$ and $mag_2$, in the top panel. In the bottom panel, we plot the residuals between the two measurements, which we call $\Delta mag$, as a function of one of the two measured magnitudes. We find slight deviation from unit slope, resulting in a median value of $\Delta mag = +0.03$, which we plot as a red dashed line in the bottom panel. We also find scatter about this median offset approximating a Gaussian distribution with a standard deviation of 0.034 mag for the central range of magnitude, excluding outliers. The brightest and faintest stars show a somewhat larger dispersion. We note that this scanning-induced error can be mitigated by taking multiple scans of the plate.
\begin{figure*}
    \centering
    \includegraphics[width=\textwidth]{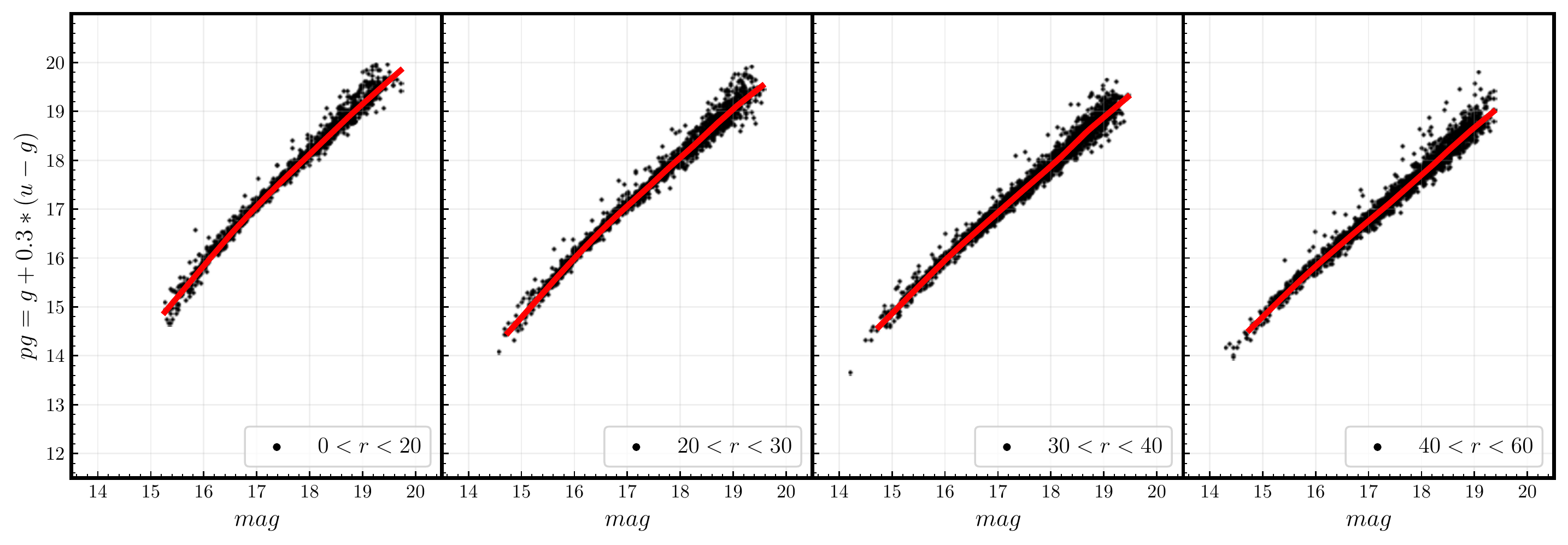}
    \caption{Aperture measurements within a four-pixel radius of stars in four bins of distance from the plate center, compared to SDSS photometry. A non-parametric LOESS curve is fit for each individual radial bin.}
    \label{fig:quad_plot}
\end{figure*}

\begin{figure*}
    \centering
    \includegraphics[width=\textwidth]{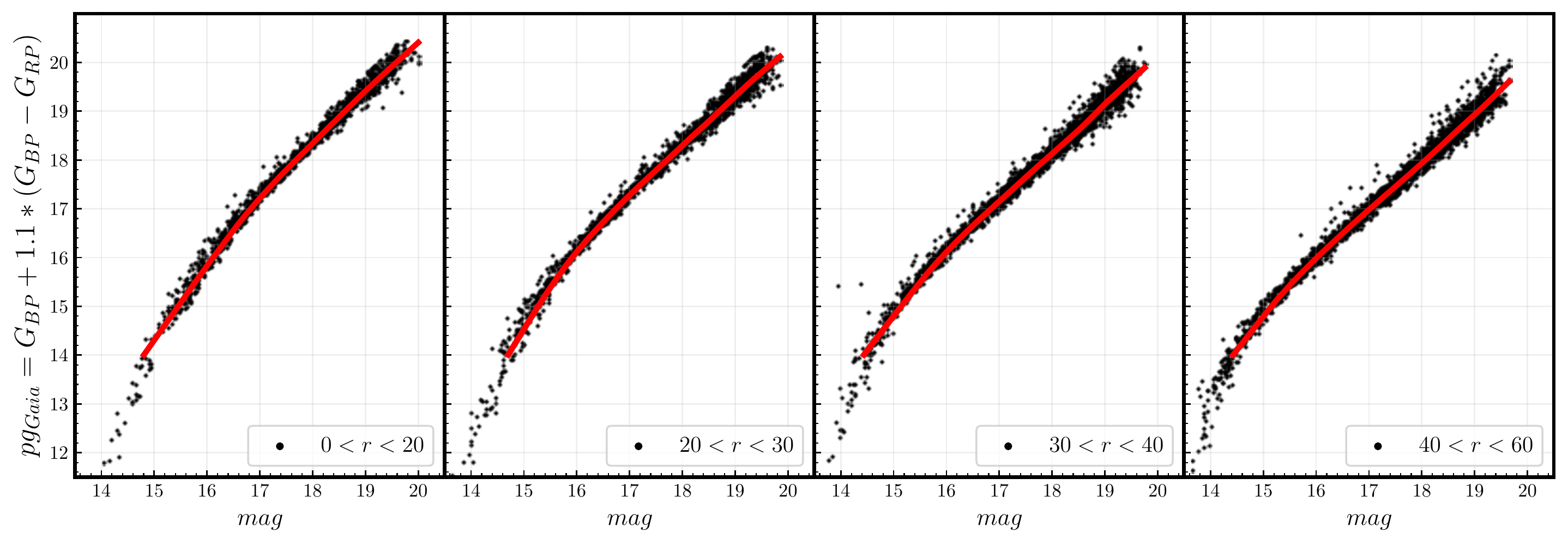}
    \caption{The same as \figref{quad_plot}, comparing the plate photometry to \Gaia, rather than SDSS. While we fit a calibration curve only to stars with $pg > 14$, there is a clear one-to-one relationship between $mag$ and the transformed \Gaia magnitude $pg$ extending to as bright as 12th magnitude.}
    \label{fig:quad_plot2}
\end{figure*}

\subsection{Photometric Calibration to SDSS and \Gaia}
\label{sec:calib}
To calibrate our stellar measurements, as already described for the astrometric comparison, we crossmatch our source catalog with  \Gaia DR2 and SDSS (independently) using \code{TOPCAT}.
\par The SDSS and \Gaia photometry differ in their utility for sake of comparison to the plate: the spectral sensitivity of the plate is not expected to be a good match to the \Gaia broadband filter bandpasses, while the SDSS $u$ and $g$ filters bracket the expected spectral sensitivity of the plate more closely. However, the \Gaia photometry extends to brighter magnitudes than the SDSS photometry after removing measurements that fail the SDSS ``clean" flag. Therefore, we proceed with a comparison of the plate photometry against both of these catalogs. \par First considering the SDSS photometry, we begin by defining a transformed SDSS magnitude called here $pg$ (shorthand for ``photographic magnitude"), according to the following equation:
$$ pg = g + 0.3*(u-g)$$
where the color term is empirically derived to minimize the trend of the residuals $pg - mag$  on $u-g$ color, where $mag$ is the average of the APT measurements from the two perpendicular scans. For comparison between $pg$ and $mag$, we add a ``zero-point`` to our measurements of $mag$ to impose that $pg = mag$ at 16th magnitude (chosen because this magnitude roughly corresponds to where the relationship experiences the least scatter, after application of the color term). 
\par We then seek a one-to-one calibration curve relating the measured plate magnitude, $mag$, to the $pg$ magnitude derived from the SDSS photometry. Since we detect a dependence of the magnitude residuals ($pg - mag$) on radial position, we consider separate intervals $r < 20$ arcmin, $20 < r < 30$, $30 < r < 40$, and $40 < r < 60$. Such a radial dependence is expected, as star images farther than about 15 arcminutes from the center of the plate are noticeably affected by coma, an aberration that increases towards the edge of the field.  Therefore, for these stars, a smaller fraction of the source's light will be included within the aperture and we can expect that the $mag$ measurements will depend on radial position on the plate, inducing radially-dependent residuals. We find that this aberration pattern does seem to be centered on the field of view, indicating the telescope was properly collimated.
\par In \figref{quad_plot}, we include a scatter plot of $pg$ vs $mag$ for each of these bins, excluding stars with residuals greater than 0.75 mag. Following a similar procedure to \citet{Tang_2013}, we fit a curve using the Locally-Weighted Scatterplot Smoothing (LOESS) fitting technique to all stars with $mag > 13.75$, depicted in red in each panel. This choice of non-parametric calibration curve is motivated by the fact that the local weighting allows for the curve fit to vary in behavior to match the trend in the relationship between $pg$ and $mag$ across the range of magnitudes. This curve was found to produce smaller residuals than some other common analytic functions, include a simple line (over the magnitude range where such an approximation is appropriate), low order polynomials, and cubic splines. 
\par In general, the relation between $pg$ and $mag$ reveals that there is a close correspondence of the plate photometry with SDSS photometry over a range of 4-5 magnitudes. The fainter stars naturally have larger errors because of lower signal-to-noise ratio, and the brighter stars are more difficult to measure because of the high densities encountered in their cores. In quantitative terms, we find typical median residuals of $\sim 0.078$ mag for stars in each bin against the LOESS curves (after removing outlier sources with residuals $> 0.75$ mag) for the sources fainter than $mag = 15$. For comparison, performing a global fit of all plate sources with $mag > 15$ results in a median residual of $\sim 0.14$ mag, reaffirming the necessity of utilizing radial bins. However, the binned non-parametric calibration approach may run a risk of overfitting on plates with fewer stars. We conclude that the median
residual of 0.078 mag suggests that the plate noise dominates over the scanner noise, the latter of which was previously found to be 0.034 mag in \secref{calib}.

In \figref{quad_plot2}, we repeat the same analysis for our plate photometry with respect to the \Gaia photometry. For the \Gaia passbands, we begin with an equation of similar form: $$ pg_{\rm Gaia} = G_{\rm BP} + 1.1*(G_{\rm BP} - G_{\rm RP})$$ 
where our measured $mag$ is scaled to match $pg_{Gaia}$ at 16th magnitude, and the color-term coefficient is derived to minimize color dependence in the residuals $pg_{Gaia} - mag$. While we find comparable residuals across a wide range of magnitudes compared to the results depicted in \figref{quad_plot}, the \Gaia DR2 catalog includes measurements of bright stars that lacked measurements in the SDSS catalog. While we do not attempt to fit our calibration curve to these brighter magnitudes, we note that the clear, low-scatter relationship between $pg_{Gaia}$ and $mag$ suggests that bright stars may still be measured well. 
\par We note that stellar photometry is more demanding than surface photometry of a diffuse source because each star image comprises a wide range of intensities and the gradients within each star image are relatively large. Nevertheless, the quality of these photometric results is consistent with the results of the surface brightness comparison in \secref{conversion}.

\subsection{Limiting Magnitude of Ry60}
\par To estimate the limiting magnitude of Ry60, we conducted a special run in APT designed to detect the faintest stars. For this run we used an aperture radius of three pixels and a (low) source detection threshold of 2.5 sigma, and we set the minimum number of pixels per source at four.  We then matched the new APT catalog with SDSS for the same region.
\par
\begin{figure}
    \centering
    \includegraphics[width=250pt]{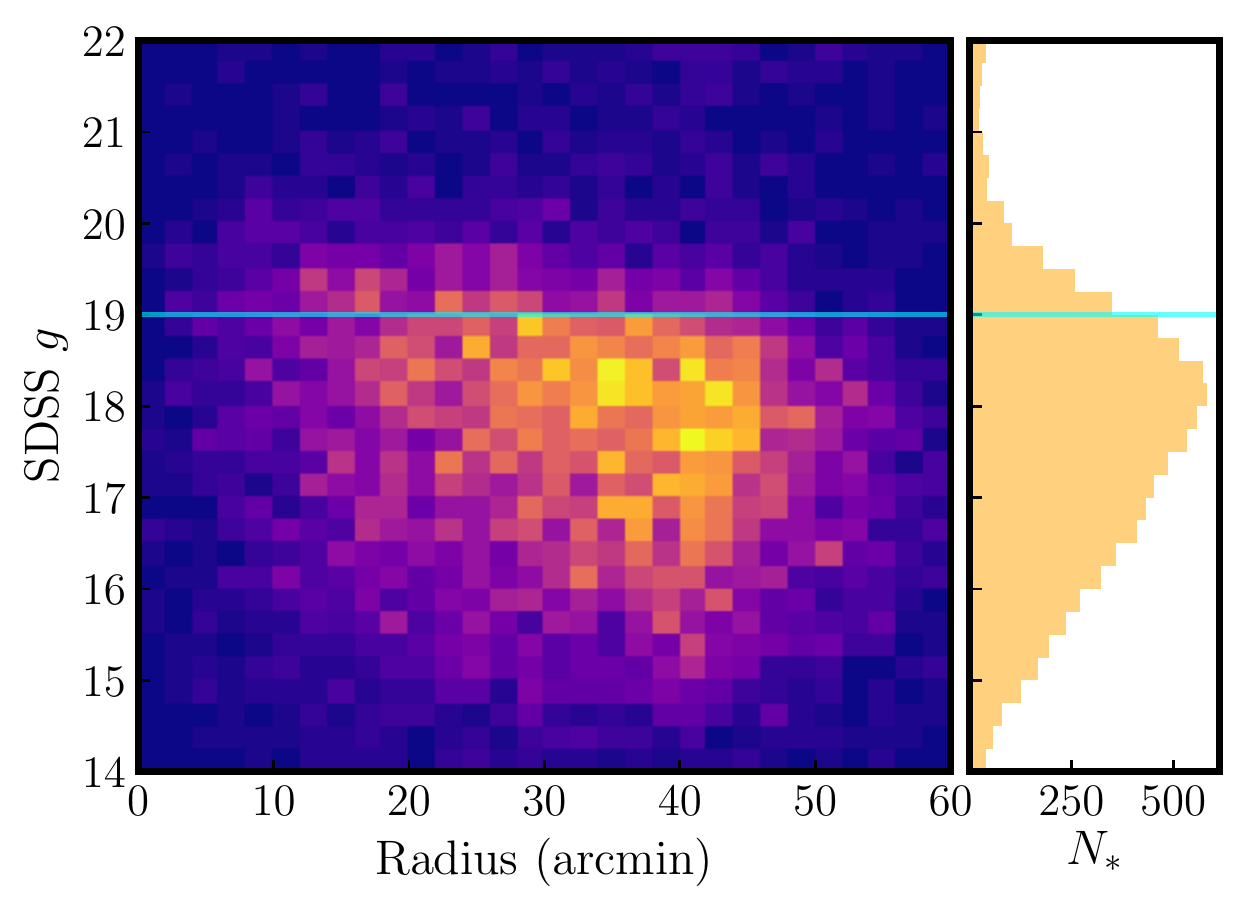}
    \caption{SDSS $g-$band magnitude vs. radius from plate center for matched stars. The limiting magnitude for the stars detected by APT is $g \lesssim 19$. The center of the field reaches  greater depths than the edges because the star images at the center are sharper and are more easily detected.  The tail to fainter magnitudes represents mismatched stars.}
    \label{fig:limitingmag}
\end{figure}

 In \figref{limitingmag}, we plot the number density of stars binned by SDSS $g-$band magnitude and radius from the plate center (in arcminutes). There is a clear dropoff in the number of stars fainter than a magnitude of SDSS $g\sim 19$, which we take to be the approximate limiting magnitude of the plate. It is evident that the depth is greater near the center of the plate due to better image quality there.
\section{Guest Star} \label{sec:guest}
\begin{figure}[h]
    \centering
    \includegraphics[width=250pt]{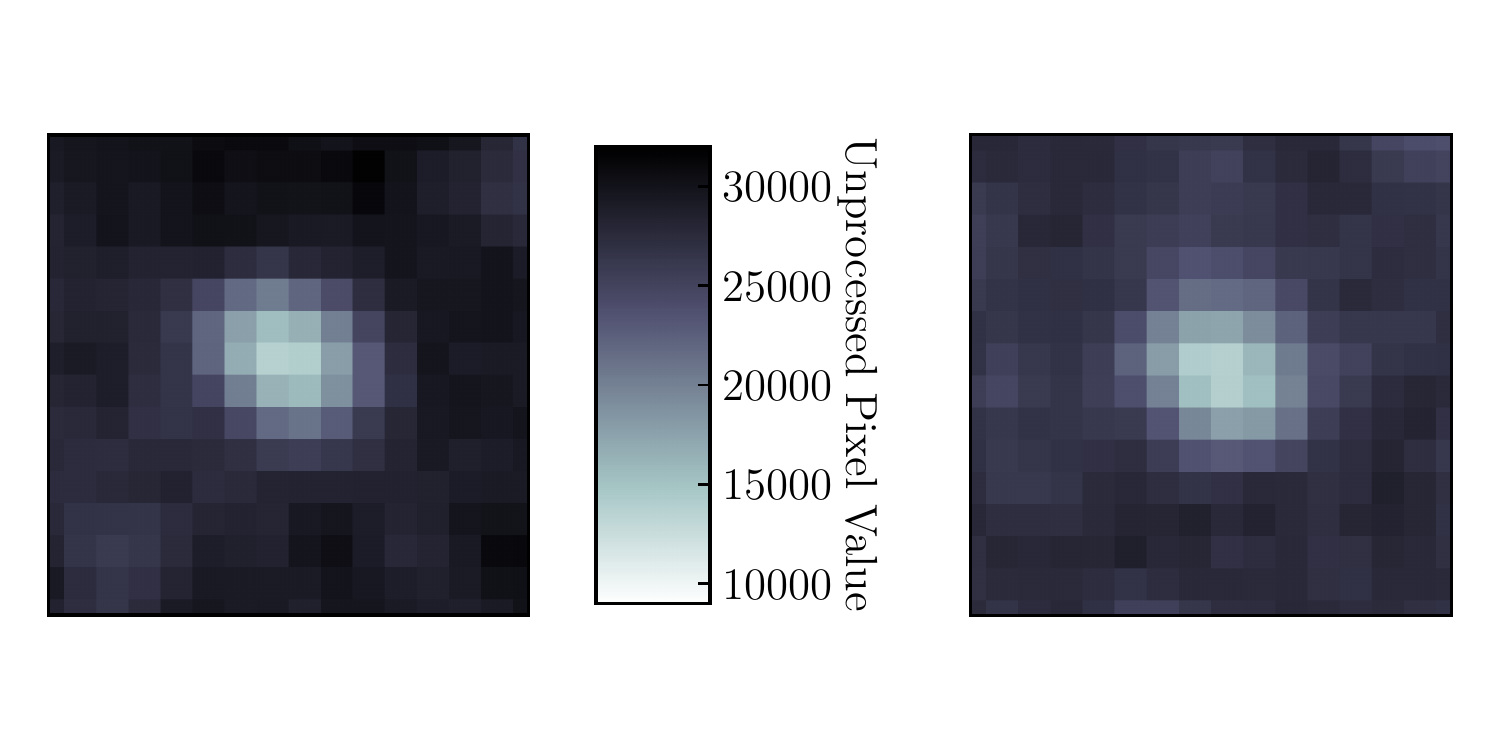}
    \caption{Comparison between the ``guest star" (left) and another star of similar magnitude which also appears in front of NGC~7331 on the plate (right). Both images are the same size and use the same color scaling (note that brighter areas have smaller scanner values due to the original, unprocessed file being a negative).}
    \label{fig:starfig}
\end{figure}

Plate Ry60 shows an unidentified, point-like source (hereafter, the ``guest star") projected onto the disk of NGC~7331, 2.9 arcminutes north of the galaxy's nucleus, near (R.A., Decl.) = ($339.2710\degree, +34.4539\degree$) - see \figref{n7331}.  There is no trace of a star at this location on the SDSS images, nor on other images of the galaxy that we inspected.  Another plate taken with the Yerkes Observatory 24-inch reflector, Ry66, has an unknown date but is likely to be of similar vintage, yet Ry66 also does not show the star. Both sides of the glass plate Ry60 have been visually inspected with a magnifier, and we have determined that the image indeed looks like a real astronomical source, as seen in \figref{starfig}, as opposed to an artifact. Therefore, under this assumption, we proceed to make a photometric measurement of the object.
\par Despite appearing against the disk of a large galaxy, the source-finding step of APT successfully found the image. To explore the sensitivity to uncertainties in the sky background, we used APT in manual mode to vary the dimensions of the annulus within which the sky background is determined, and we also explored what happens if the mean of the pixels within the annulus is used to estimate the sky level, as opposed to the median. All of these experiments led to variations at the level of less than about 0.03 mag and we conclude that the measurement is not strongly affected by the galaxy per se, nor to irregularities in the background disk of the galaxy. 
\par With this $mag$ measurement, we then utilized the calibration curves fit in \secref{photometry} to transform this measurement to the $pg$ system (based on the SDSS photometry). Doing so, we find $pg = 16.60$ mag for the object on August 19, 1903 (MJD 16345). Based on the median residuals of our stellar measurements depicted in \figref{quad_plot}, we estimate the error in this measurement to be $\pm 0.08$ mag.

\par The ``guest star" cannot be a classical nova in the disk of NGC~7331, which is located at a distance of  $\sim 14.5$ Mpc \citep{Freedman}, as it would be much too faint to be observable by a small-aperture telescope with photography, even in three hours of integration time. It is unlikely to be a Solar System object as the image is not trailed in the three-hour exposure (the field is at an ecliptic latitude of $\roughly 39 \degree$). Therefore, informed by our apparent magnitude measurement, we find that the star is likely to be a previously unreported supernova. 
\par Our measurement from the Ry60 plate can be most directly compared to the measured magnitudes of three supernovae already known to have occurred in NGC~7331. The only one of these events with publicly reported $g-$band measurements was the Type II supernova SN2013bu, which was found have a peak apparent magnitude of $g_{\rm max} \sim 14.0$ \citep{SN2013bu}. While further comparison is not possible based upon a single measurement epoch for the plate's ``guest star," the approximate consistency of our measurement with this modern observation supports the plausibility of this newly discovered object as a candidate supernova.

\section{Conclusion}
\label{sec:conclusion}
We have demonstrated that a commercial flatbed scanner is capable of extracting scientifically useful astrometric and photometric data from photographic plates. The scanner that we tested has quasi-periodic positional error in the scan direction with an amplitude of about 1.5 arcseconds, a feature also noted by \citet{Vicente07} and \citet{Pakuliak}. However this error is generally small enough to permit reliable identifications of the stars for photometric work. We found that the response of the scanner to transmitted light requires an expanded scale, but when that is calibrated out, the data faithfully reproduce values over a density and log intensity range approaching 2.0. Independent scans reduced for stellar photometry are consistent at the few-percent level, and stars of intermediate brightness near the center of the field are measured for brightness to better than 8 percent based on matches to \Gaia and SDSS photometry, a result that is consistent with \citet{Sokolovsky}. 
\par Automation of the procedures we have developed renders data accessible at low cost from astronomical photographic plate collections,  at least in principle.  Such a capability would of course be invaluable for historical investigation of variable stars, AGN, and other transients \citep[e.g,][]{Grindlay14}. To make the files navigable and thus immediately useful to researchers requires not only automation, but also access to digital versions of the logbooks and metadata associated with the plates.  We intend to continue our development program with this long-term goal in mind.

\section{Acknowledgments}
We are grateful for continuing access to the plate material housed at Yerkes Observatory, now operated by the Yerkes Future Foundation. Travel to Yerkes Observatory and other expenses have been generously supported by the Physical Sciences Collegiate Division of the University of Chicago. \par We have used the resources of the Regenstein Library Special Collections Research Center to locate some observing logbooks. The scanner we used was purchased for this project thanks to the Kathleen and Howard Zar Science Library Fund.  
\par We have benefited from continuing engagement with the University of Chicago Library's Preservation Department with their experience with digitization techniques, especially Christina Miranda-Izguerra, Lauren Reese, and Sherry Byrne. General explorations of the Yerkes plate vault, associated documentation such as the logbooks, and the background history of Yerkes and Yerkes astronomers have also been in close coordination with others at the University of Chicago Library, especially Barbara Kern, Elisabeth Long, Andrea Twiss-Brooks, and Clara del Junco. We also thank Wayne Osborn for his help with using the logbooks to locate plates in the collection.
\par This work has made use of data from the European Space Agency (ESA) mission {\it Gaia} (\url{https://www.cosmos.esa.int/gaia}), processed by the {\it Gaia} Data Processing and Analysis Consortium (DPAC, \url{https://www.cosmos.esa.int/web/gaia/dpac/consortium}). Funding for the DPAC
has been provided by national institutions, in particular the institutions
participating in the {\it Gaia} Multilateral Agreement.  
\par This work has also made use of the SDSS DR7 \citep{sdssDR7}. Funding for the SDSS and SDSS-II has been provided by the Alfred P. Sloan Foundation, the Participating Institutions, the National Science Foundation, the U.S. Department of Energy, the National Aeronautics and Space Administration, the Japanese Monbukagakusho, the Max Planck Society, and the Higher Education Funding Council for England. The SDSS Web Site is http://www.sdss.org/.
\par The SDSS is managed by the Astrophysical Research Consortium for the Participating Institutions. The Participating Institutions are the American Museum of Natural History, Astrophysical Institute Potsdam, University of Basel, University of Cambridge, Case Western Reserve University, University of Chicago, Drexel University, Fermilab, the Institute for Advanced Study, the Japan Participation Group, Johns Hopkins University, the Joint Institute for Nuclear Astrophysics, the Kavli Institute for Particle Astrophysics and Cosmology, the Korean Scientist Group, the Chinese Academy of Sciences (LAMOST), Los Alamos National Laboratory, the Max-Planck-Institute for Astronomy (MPIA), the Max-Planck-Institute for Astrophysics (MPA), New Mexico State University, Ohio State University, University of Pittsburgh, University of Portsmouth, Princeton University, the United States Naval Observatory, and the University of Washington.

\facility{\Gaia.}
\software{\code{astropy} \citep{astropy:2013,astropy:2018}, \code{Matplotlib} \citep{Hunter:2007}, \code{numpy} \citep{numpy:2011}}, \code{TOPCAT} \citep{topcat}, \code{Aperture Photometry Tool} \citep{APT}, \code{scipy} \citep{scipy}

\bibliography{main}

\end{document}